\documentclass[twocolumn,letterpaper]{IEEEAerospace3}
\usepackage[]{
amssymb,amsfonts,amsmath,amstext,times,epsfig}

\newcommand{\ignore}[1]{} %%% {} empty inside = %% comment

\ignore{
}

\begin{document}
%paper title
\title{The Science, Technology and Mission Design for \\ the Laser Astrometric Test of Relativity}

%author's name, address, etc.
\author{
Slava G.\ Turyshev\\
Jet Propulsion Laboratory, 
California Institute of Technology, \\
4800 Oak Grove Drive, Pasadena, CA 91109, USA\\
turyshev@jpl.nasa.gov\\
\thanks{{\footnotesize 
0-7803-9546-8/06/\$20.00 \copyright~2006 IEEE}}
\thanks{{\footnotesize IEEEAC paper \# XXX}} %% creates required footnotes with copyright and paper #
}

\ignore{ %%% can be used for your purposes, not in the submitted version
\markboth{IEEE Aerospace Conference 2006}
{Author: Title}
}

\maketitle

\pagestyle{plain}

\begin{abstract}
%%%%% 150 words Abstract
The Laser Astrometric Test of Relativity (LATOR) is a  Michelson-Morley-type experiment designed to test the Einstein's general theory of relativity in the most intense gravitational environment available in the solar system -- the close proximity to the Sun. By using independent time-series of highly accurate measurements of the Shapiro time-delay (laser ranging accurate to 1 cm) and interferometric astrometry (accurate to 0.1 picoradian), LATOR will measure gravitational deflection of light by the solar gravity with accuracy of 1 part in a billion, a factor $\sim$30,000 better than currently available.  LATOR will perform series of highly-accurate tests of gravitation and cosmology in its search for cosmological remnants of scalar field in the solar system.   We present science, technology and mission design for the LATOR mission.

\end{abstract}

%% Allows to create Keywords if it is needed (not for the submitted version)
\ignore{
}

\tableofcontents

%*********************1) INTRODUCTION
\section{Introduction}
\label{sec:intro}

The Laser Astrometric Test of Relativity (LATOR) is a space-based experiment that is designed to significantly improve the tests of relativistic gravity in the solar system. If the Eddington's 1919 experiment was performed to confirm the Einstein's general theory of relativity, LATOR is motivated to search for physics beyond the Einstein's theory of gravity with an unprecedented accuracy \cite{lator_cqg_2004}.  In fact, this mission is designed to address the questions of fundamental importance to modern physics by searching for a cosmologically-evolved scalar field that is predicted by modern theories of gravity and cosmology, and also by superstring and brane-world models. LATOR will also test the cosmologically motivated theories that attempt to explain the small acceleration rate of the Universe (so-called `dark energy') via modification of gravity at very large, horizon or super-horizon distances. 

By studying the effect of gravity on light and measuring the Eddington parameter $\gamma$, this mission will tests the presently viable alternative theories of gravity, namely the scalar-tensor theories. The value of the parameter $\gamma$ may hold the key to the solution of the most fundamental questions concerning the evolution of the universe.  In the low energy approximation suitable for the solar system, a number of modern theories of gravity and cosmology studied as methods for gravity quantization or proposed as an explanation to the recent cosmological puzzles, predict measurable contributions to the parameter $\gamma$ at the level of $(1-\gamma)\sim 10^{-6}-10^{-8}$; detecting this deviation is LATOR's primary objective.  With the accuracy of one part in $10^{9}$, this mission could discover a violation or extension of general relativity, and/or reveal the presence of any additional long range interaction. 

The LATOR test will be performed in the solar gravity field using optical interferometry between two micro-spacecraft.  Precise measurements of the angular position of the spacecraft will be made using a fiber coupled multi-chanelled optical interferometer on the ISS with a 100 m baseline. The primary objective of the LATOR mission will be to measure the gravitational deflection of light by the solar gravity to accuracy of 0.1 picoradians (prad), which corresponds to $\sim$10 picometers (pm) on a 100 m interferometric baseline. A combination of laser ranging among the spacecraft and direct interferometric measurements will allow LATOR to measure deflection of light in the solar gravity by a factor of more than 30,000 better than had recently been accomplished with the Cassini spacecraft. In particular, this mission will not only measure the key PPN parameter $\gamma$ to unprecedented levels of accuracy of one part in 10$^9$; it will also reach ability to measure the next post-Newtonian order ($\propto G^2$) of light deflection resulting from gravity's intrinsic non-linearity and a number of corresponding effects to very significant level of accuracy (discussed below). LATOR is envisaged as a partnership between US and Europe with clear areas of responsibility between the corresponding space agencies: NASA provides the deep space mission components, while access to and optical infrastructure on the ISS would be an ESA contribution. 

This paper is organized as follows: Section \ref{sec:sci_mot} discusses the theoretical framework and science motivation for the precision gravity tests in the solar system; it also presents the science objectives for the LATOR experiment. Section \ref{sec:lator_description} provides an overview for the LATOR experiment including basic elements of the current mission and optical designs. 
Section~\ref{sec:interferometry} addresses design of the LATOR optical receivers and the long-baseline optical interferometer. 
Section~\ref{sec:conc} presents conclusions and discusses the next steps that will be taken in the development of LATOR.  

%*********************1) INTRODUCTION
\section{Scientific Motivation}
\label{sec:sci_mot}

The recent remarkable progress in observational cosmology has again submitted general relativity to a test by suggesting a non-Einsteinian model of universe's evolution. From the theoretical standpoint, the challenge is even stronger---if the gravitational field is to be quantized, the general theory of relativity will have to be modified. This is why the recent advances in the scalar-tensor extensions of gravity, that are consistent with the current inflationary model of the Big Bang, have motivated new search for a very small deviation of from Einstein's theory, at the level of three to five orders of magnitude below the level tested by experiment. 

In this section we will consider the recent motivations for the high-accuracy gravitational tests. We will also present the scientific objectives of the LATOR experiment.

\subsection{The PPN Formalism}

Generalizing on a phenomenological parameterization of the gravitational metric tensor field which Eddington originally developed for a special case, a method called the parameterized post-Newtonian (PPN) metric has been developed \cite{Ken_EqPrinciple68b,Ken_LLR68,Ken_2PPN_87,WillNordtvedt72,Will_book93}.
This method  represents the gravity tensor's potentials for slowly moving bodies and weak interbody gravity, and it is valid for a broad class of metric theories including general relativity as a unique case.  The several parameters in the PPN metric expansion vary from theory to theory, and they are individually associated with various symmetries and invariance properties of underlying theory.  Gravity experiments can be analyzed in terms of the PPN metric, and an ensemble of experiments will determine the unique value for these parameters, and hence the metric field, itself.

In locally Lorentz-invariant theories the expansion of the metric field for a single, slowly-rotating gravitational source in PPN coordinates is given by:
\begin{eqnarray}
\label{eq:metric}
g_{00}\hskip -6pt &=&\hskip -5pt 1-2\frac{M}{r}Q(r,\theta) +2\beta\frac{M^2}{r^2}+{\cal O}(c^{-6}),\nonumber\\ 
g_{0i}\hskip -6pt &=&\hskip -5pt  2(\gamma+1)\frac{[\vec{J}\times \vec{r}]_i}{r^3}+
{\cal O}(c^{-5}),\\ 
g_{ij}\hskip -6pt &=&\hskip -9pt -~\delta_{ij}\Big[1+
2\gamma \frac{M}{r} Q(r,\theta)+
\frac{3}{2}\delta \frac{M^2}{r^2}\Big]
\hskip -1pt +\hskip -0pt {\cal O}(c^{-6}),\nonumber
\end{eqnarray}

\noindent where $M$ and $\vec J$ being the mass and angular momentum of the Sun, $Q(r,\theta)=1-J_2\frac{R^2}{r^2}\frac{3\cos^2\theta-1}{2}$, with $J_2$ being the quadrupole moment of the Sun and $R$ being its radius.  $r$ is the distance between the observer and the center of the Sun.  $\beta, \gamma, \delta$ are the PPN parameters and in general relativity they are all equal to $1$. The $M/r$ term in the $g_{00}$ equation is the Newtonian limit; the terms multiplied by the post-Newtonian parameters $\beta, \gamma$,  are post-Newtonian terms. The term multiplied by the post-post-Newtonian parameter  $\delta$ also enters the calculation of the relativistic light deflection \cite{Ken_cqg96}.

This PPN expansion serves as a useful framework to test relativistic gravitation in the context of the LATOR mission. In the special case, when only two PPN parameters ($\gamma$, $\beta$) are considered, these parameters have clear physical meaning. Parameter $\gamma$  represents the measure of the curvature of the space-time created by a unit rest mass; parameter  $\beta$ is a measure of the non-linearity of the law of superposition of the gravitational fields in the theory of gravity. General relativity, which corresponds to  $\gamma = \beta$  = 1, is thus embedded in a two-dimensional space of theories. The Brans-Dicke is the best known theory among the alternative theories of gravity.  It contains, besides the metric tensor, a scalar field and an arbitrary coupling constant $\omega$, which yields the two PPN parameter values $\gamma = (1+ \omega)/(2+ \omega)$, and $\beta$  = 1.  More general scalar tensor theories yield values of $\beta$ different from one.

The most precise value for the PPN parameter  $\gamma$ is at present given by the Cassini mission \cite{cassini_ber} as: $\gamma -1 = (2.1\pm2.3)\times10^{-5}$. Using the recent Cassini result \cite{cassini_ber} on the PPN  $\gamma$, the parameter $\beta$ was measured as $\beta-1=(0.9\pm1.1)\times 10^{-4}$ from LLR \cite{Williams_etal_2004,LLR_beta_2004}. The next order PPN parameter $\delta$ has not yet been measured though its value can be inferred from other measurements.

The Eddington parameter $\gamma$, whose value in general relativity is unity, is perhaps the most fundamental PPN parameter, in that $\frac{1}{2}(1-\gamma)$ is a measure, for example, of the fractional strength of the scalar gravity interaction in scalar-tensor theories of gravity \cite{Damour_EFarese96a,Damour_EFarese96b}.  Within perturbation theory for such theories, all other PPN parameters to all relativistic orders collapse to their general relativistic values in proportion to $\frac{1}{2}(1-\gamma)$. This is why measurement of the first order light deflection effect at the level of accuracy comparable with the second-order contribution would provide the crucial information separating alternative scalar-tensor theories of gravity from general relativity \cite{Ken_2PPN_87} and also to probe possible ways for gravity quantization and to test modern theories of cosmological evolution \cite{Damour_Nordtvedt_93a,Damour_Nordtvedt_93b,DamourPolyakov94a,DamourPolyakov94b,DPV02a,DPV02b}. The LATOR mission is designed to directly address this issue with an unprecedented accuracy.

Over the recent decade, the technology has advanced to the point that one can consider carrying out direct tests in a weak field to second order in the field strength parameter ($\propto G^2$). Although any measured anomalies in first or second order metric gravity potentials will not determine strong field gravity, they would signal that modifications in the strong field domain will exist.  The converse is perhaps more interesting:  if to high precision no anomalies are found in the lowest order metric potentials, and this is reinforced by finding no anomalies at the next order, then it follows that any anomalies in the strong gravity environment are correspondingly quenched under all but exceptional circumstances.\footnote{For example, a mechanism of a ``spontaneous-scalarization'' that, under certain circumstances, may exist in tensor-scalar theories \cite{Damour_EFarese93}.} 

We shall now discuss the recent motivations for the precision gravity experiments.

\subsection{Motivations for Precision Gravity Experiments}
\label{sec:mot} 

The continued inability to merge gravity with quantum mechanics, and recent cosmological observations indicate that the pure tensor gravity of general relativity needs modification.  The tensor-scalar theories of gravity, where the usual general relativity tensor field coexists with one or several long-range scalar fields, are believed to be the most promising extension of the theoretical foundation of modern gravitational theory. The superstring, many-dimensional Kaluza-Klein and inflationary cosmology theories have revived interest in the so-called ``dilaton fields,'' i.e. neutral scalar fields whose background values determine the strength of the coupling constants in the effective four-dimensional theory.  The importance of such theories is that they provide a possible route to the quantization of gravity and the unification of physical laws. 

Although the scalar fields naturally appear in the theory, their inclusion predicts different relativistic corrections to Newtonian motions in gravitating systems. These deviations from general relativity lead to a violation of the Equivalence Principle (either weak or strong or both), modification of large-scale gravitational phenomena, and generally lead to space and time variation of physical ``constants.'' As a result, this progress has provided new strong motivation for high precision relativistic gravity tests.

\subsubsection{Tensor-Scalar Theories of Gravity}
\label{sec:mot_theories}

Recent theoretical findings suggest that the present agreement between general relativity and experiment might be naturally compatible with the existence of a scalar contribution to gravity. In particular, Damour and Nordtvedt \cite{Damour_Nordtvedt_93a,Damour_Nordtvedt_93b} (see also \cite{DamourPolyakov94a,DamourPolyakov94b} for non-metric versions of this mechanism together with \cite{DPV02a,DPV02b} for the recent summary of a dilaton-runaway scenario) have found that a scalar-tensor theory of gravity may contain a ``built-in'' cosmological attractor mechanism toward general relativity.  These scenarios assume that the scalar coupling parameter $\frac{1}{2}(1-\gamma)$ was of order one in the early universe (say, before inflation), and show that it then evolves to be close to, but not exactly equal to, zero at the present time \cite{lator_cqg_2004}. 

Under some assumptions (see e.g. \cite{Damour_Nordtvedt_93a,Damour_Nordtvedt_93b}) one can even estimate what is the likely order of magnitude of the left-over coupling strength at present time which, depending on the total mass density of the universe, can be given as $1-\gamma \sim 7.3 \times 10^{-7}(H_0/\Omega_0^3)^{1/2}$, where $\Omega_0$ is the ratio of the current density to the closure density and $H_0$ is the Hubble constant in units of 100 km/sec/Mpc. Compared to the cosmological constant, these scalar field models are consistent with the supernovae observations for a lower matter density, $\Omega_0\sim 0.2$, and a higher age, $(H_0 t_0) \approx 1$. If this is indeed the case, the level $(1-\gamma) \sim 10^{-6}-10^{-7}$ would be the lower bound for the present value of PPN parameter $\gamma$ \cite{Damour_Nordtvedt_93a,Damour_Nordtvedt_93b}. 

More recently, \cite{DPV02a,DPV02b} have estimated $\frac{1}{2}(1-\gamma)$, within the framework compatible with string theory and modern cosmology, which basically confirms the previous result \cite{Damour_Nordtvedt_93a,Damour_Nordtvedt_93b}. This recent analysis discusses a scenario when a composition-independent coupling of dilaton to hadronic matter produces detectable deviations from general relativity in high-accuracy light deflection experiments in the solar system. This work assumes only some general property of the coupling functions (for large values of the field, i.e. for an ``attractor at infinity'') and then only assume that $(1-\gamma)$ is of order of one at the beginning of the controllably classical part of inflation.
It was shown  in \cite{DPV02b} that one can relate the present value of $\frac{1}{2}(1-\gamma)$ to the cosmological density fluctuations.
For the simplest inflationary potentials (favored by WMAP mission, i.e. $m^2 \chi^2$ \cite{[4c]}) \cite{DPV02a,DPV02b} found that the present value of $(1-\gamma)$ could be just below $10^{-7}$.
In particular, within this framework $\frac{1}{2}(1-\gamma)\simeq\alpha^2_{\rm had}$, where $\alpha_{\rm had}$ is the dilaton coupling to hadronic matter; its value depends on the model taken for the inflation potential $V(\chi)\propto\chi^n$, with $\chi$ being the inflation field; the level of the expected deviations from general relativity is $\sim0.5\times10^{-7}$  for $n = 2$ \cite{DPV02b}. Note that these predictions are based on the work in scalar-tensor extensions of gravity which are consistent with, and indeed often part of, present cosmological models.

The analyses discussed above not only motivate new searches for very small deviations of relativistic gravity in the solar system, they also predict that such deviations are currently present in the range from $10^{-5}$ to $5\times10^{-8}$ for $\frac{1}{2}(1-\gamma)$, i.e. for observable post-Newtonian deviations from general relativity predictions and, thus, should be easily detectable with LATOR. This would require measurement of the effects of the next post-Newtonian order ($\propto G^2$) of light deflection resulting from gravity's intrinsic non-linearity. An ability to measure the first order light deflection term at the accuracy comparable with the effects of the second order is of the utmost importance for gravitational theory and a major challenge for the 21st century fundamental physics.  

\subsubsection{Experimental Motivations for High-Accuracy Gravity Tests}
\label{sec:mot_experiment}

Recent astrophysical measurements of the angular structure of the cosmic microwave background \cite{deBernardis_CMB2000}, the masses of large-scale structures \cite{Peacock_LargeScale01}, and the luminosity distances of type Ia supernovae \cite{perlmutter99,Riess_supernovae98} have placed stringent constraints on the cosmological constant $\Lambda$ and also have led to a revolutionary conclusion: the expansion of the universe is accelerating. The implication of these observations for cosmological models is that a classically evolving scalar field currently dominates the energy density of the universe. Such models have been shown to share the advantages of  $\Lambda$:  compatibility with the spatial flatness predicted inflation; a universe older than the standard Einstein-de Sitter model; and, combined with cold dark matter, predictions for large-scale structure formation in good agreement with data from galaxy surveys. Combined with the fact that scalar field models imprint distinctive signature on the cosmic microwave background (CMB) anisotropy, they remain currently viable and should be testable in the near future. This completely unexpected discovery demonstrates the importance of testing the important ideas about the nature of gravity. 

There is now multiple evidence indicating that 70\% of the critical density of the universe is in the form of a ``negative-pressure'' dark energy component; there is no understanding as to its origin and nature. The fact that the expansion of the universe is currently undergoing a period of acceleration now seems rather well tested: it is directly measured from the light-curves of several hundred type Ia supernovae \cite{perlmutter99,Riess_supernovae98,[3c]}, and independently inferred from observations of CMB by the WMAP satellite \cite{[4c]} and other CMB experiments \cite{[6c],[5c]}. Cosmic speed-up can be accommodated within general relativity by invoking a mysterious cosmic fluid with large negative pressure, dubbed dark energy. The simplest possibility for dark energy is a cosmological constant; unfortunately, the smallest estimates for its value are 55 orders of magnitude too large (for reviews see \cite{Carroll_01,PeeblesRatra03}). Most of the theoretical studies operate in the shadow of the cosmological constant problem, the most embarrassing hierarchy problem in physics. This fact has motivated a host of other possibilities, most of which assume $\Lambda=0$, with the dynamical dark energy being associated with a new scalar field (see \cite{[carroll]} and references therein). However, none of these suggestions is compelling and most have serious drawbacks. 

Given the challenge of this problem, a number of authors considered the possibility that cosmic acceleration is not due to some kind of stuff, but rather arises from new gravitational physics (see discussion in 
\cite{Carroll_01,PeeblesRatra03,[carroll],Carroll_HT_03}).
In particular, some extensions to general relativity in a low energy regime \cite{[carroll]} were shown to predict an experimentally consistent universe evolution  without the need for dark energy. These dynamical models are expected to produce measurable contribution to the parameter $\gamma$  in experiments conducted in the solar system also at the level of $1-\gamma \sim 10^{-7}-5\times10^{-9}$, thus further motivating the relativistic gravity research. Therefore, the PPN parameter $\gamma$ may be the only key parameter that holds the answer to most of the questions discussed above. Also an anomalous parameter $\delta$ will most likely be accompanied by a `$\gamma$ mass' of the Sun which differs from the gravitational mass of the Sun and therefore will show up as anomalous $\gamma$ (see discussion in \cite{Ken_LLR_PPNprobe03}).

In summary, there are a number of theoretical and experimental reasons to question the validity of general relativity.   
The LATOR mission is designed to address theses challenges.
We shall now discuss the LATOR mission in more details.

\section{Overview of LATOR}
\label{sec:lator_description}

\begin{figure*}[t!]
 \begin{center}
\noindent  
\psfig{figure=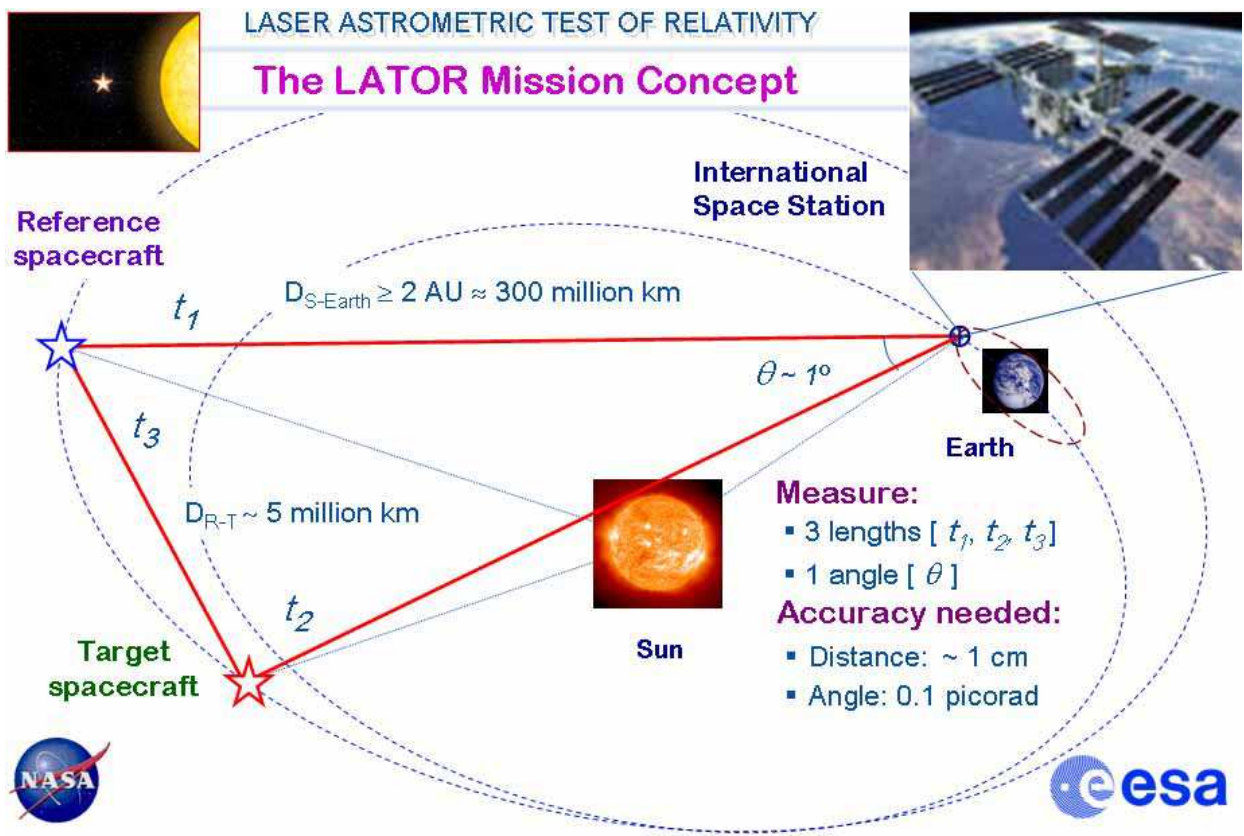,width=164mm}
\end{center}
\vskip -10pt 
  \caption{The overall geometry of the LATOR experiment.} 
\label{fig:lator}
\vskip -5pt 
\end{figure*}

The LATOR experiment uses the standard technique of time-of-fight laser ranging between two micro-spacecraft whose lines of sight pass close by the Sun and also a long-baseline stellar optical interferometer (placed above the Earth's atmosphere) to accurately measure deflection of light by the solar gravitational field in the extreme proximity to the Sun \cite{lator_cqg_2004}.  Figure \ref{fig:lator} shows the general concept for the LATOR missions including the mission-related geometry, experiment details  and required accuracies.

%************
%**************

\subsection{Science with LATOR}
\label{sec:lator_science}

LATOR is a Michelson-Morley-type experiment designed to test the pure tensor metric nature of gravitation -- a fundamental postulate of Einstein's theory of general relativity \cite{lator_cqg_2004}.  With its focus on gravity's action on light propagation it complements other tests which rely on the gravitational dynamics of bodies.  The idea behind this experiment is to use a combination of independent time-series of highly accurate measurements of the gravitational deflection of light in the immediate proximity to the Sun along with measurements of the Shapiro time delay on the interplanetary scales (to a precision respectively better than $10^{-13}$~rad and 1 cm).  Such a combination of observables is unique and enables LATOR to significantly improve tests of relativistic gravity.

The LATOR's primary mission objective is to measure the key post-Newtonian Eddington parameter $\gamma$ with an accuracy of a part in 10$^9$.  
When the light deflection in solar gravity is concerned, the magnitude of the first order effect as predicted by general relativity for the light ray just grazing the limb of the Sun is $\sim1.75$ arcsecond (asec). (Note that 1 arcsec $\simeq5~\mu$rad; when convenient, below we will use the units of radians and arcseconds interchangeably.) The effect varies inversely with the impact parameter. The second order term is almost six orders of magnitude smaller resulting in  $\sim 3.5$ microarcseconds ($\mu$as) light deflection effect, and which falls off inversely as the square of the light ray's impact parameter \cite{lator_cqg_2004,Ken_2PPN_87,epstein_shapiro_80,FishbachFreeman80,RichterMatzner82a,RichterMatzner82b,RichterMatzner83}. The relativistic frame-dragging term\footnote{Gravitomagnetic frame dragging is the effect in which both the orientation and trajectory of objects in orbit around a body are altered by the gravity of the body's rotation.  It was studied by Lense and Thirring in 1918.} is $\pm 0.7 ~\mu$as, and contribution of the solar quadrupole moment, $J_2$, is sized as 0.2 $\mu$as (using theoretical value of the solar quadrupole moment $J_2\simeq10^{-7}$). The small magnitudes of the effects emphasize the fact that, among the four forces of nature, gravitation is the weakest interaction; it acts at very long distances and controls the large-scale structure of the universe, thus, making the precision tests of gravity a very challenging task.

%%********************Start Table******************
\def\reff{\vskip 4pt \par \hangindent 12pt \noindent}
\begin{table}[t!]
\caption{LATOR Mission Summary: Science Objectives 
\label{tab:summ_science}}

{\it Qualitative Objectives: }

\reff $\bullet$\hskip6pt
To test the metric nature of the Einstein's general theory of relativity in the most intense gravitational environment available in the solar system -- the extreme proximity to the Sun

\reff $\bullet$\hskip6pt
To test alternative theories of gravity and cosmology, notably scalar-tensor theories, by searching for cosmological remnants of scalar field in the solar system

\reff $\bullet$\hskip6pt
To verify the models of light propagation and motion of the gravitationally-bounded systems at the second post-Newtonian order (i.e. including effects $\propto G^2$)\\[4pt] 

{\it Quantitative Objectives: }

\reff $\bullet$\hskip6pt
To measure the key Eddington PPN parameter $\gamma$ with accuracy of 1 part in 10$^{9}$ -- a factor of 30,000 improvement in the tests of gravitational deflection of light

\reff $\bullet$\hskip6pt
To provide direct and independent measurement of the Eddington PPN parameter $\beta$ via gravity effect on light to $\sim0.01$\% accuracy

\reff $\bullet$\hskip6pt
To measure effect of the 2-nd order gravitational deflection of light with accuracy of $\sim1\times 10^{-4}$, including first ever measurement of the post-PPN parameter $\delta$ 

\reff $\bullet$\hskip6pt
To directly measure the frame dragging effect on light (first such observation and also first direct measurement of solar spin) with $\sim 1\times10^{-3}$ accuracy

\reff $\bullet$\hskip6pt
To measure the solar quadrupole moment $J_2$ (using the theoretical value of $J_2 \simeq 10^{-7}$) to 1 part in 200, currently unavailable. 
\vskip-10pt
\end{table}
 
%********************End Table******************

The first order effect of light deflection in the solar gravity caused by the solar mass monopole is $\alpha_1=1.75$ arcseconds \cite{lator_cqg_2004}, which corresponds to an interferometric delay of $d\simeq b\alpha_1\approx0.85$~mm on a $b=100$~m baseline. Using laser interferometry, we currently are able to measure distances with an accuracy (not just precision but accuracy) of $\leq$~1~pm. In principle, the 0.85 mm gravitational delay can be measured with $10^{-10}$ accuracy versus $10^{-4}$ available with current techniques. However, we use a conservative estimate for the delay of 5 pm which would produce the measurement of $\gamma$ to accuracy of 1 part in $10^{9}$ (rather than 1 part in $10^{-10}$), which would be already a factor of 30,000 accuracy improvement when compared to the recent Cassini result \cite{cassini_ber}. Furthermore, we have targeted an overall measurement accuracy of 5 pm per measurement, which for $b=100$~m this translates to the accuracy of 0.1 prad $\simeq 0.01~\mu$as. With 4 measurements per observation, this yields an accuracy of $\sim5.8\times 10^{-9}$ for the first order term. The second order light deflection is approximately 1700 pm and, with 5 pm accuracy and $\sim~400$ independent data points, it could be measured with accuracy of $\sim1$ part in $10^{4}$, including first ever measurement of the PPN parameter $\delta$.  The frame dragging effect would be measured with $\sim 1$ part in $10^{3}$ accuracy and the solar quadrupole moment can be modestly measured to 1 part in 200, all with respectable signal to noise ratios. A summary of the mission objectives are presented in Table~\ref{tab:summ_science}. Recent covariance studies performed for the LATOR mission \cite{hellings_2005,Ken_lator05} confirm the design performance parameters and also provide valuable recommendations for further mission developments. 

We shall now consider the LATOR mission architecture.

\subsection{Mission Design: Evolving Light Triangle}
\label{sec:triangle}

The LATOR mission architecture uses an evolving light triangle formed by laser ranging between two spacecraft (placed in $\sim$1 AU heliocentric orbits) and a laser transceiver terminal on the International Space Station (ISS), via European collaboration.  The objective is to measure the gravitational deflection of laser light as it passes in extreme proximity to the Sun (see Figure \ref{fig:lator}).  To that extent, the long-baseline ($\sim$100 m) fiber-coupled optical interferometer on the ISS will perform differential astrometric measurements of the laser light sources on the two spacecraft as their lines-of-sight pass behind the Sun.  As seen from the Earth, the two spacecraft will be separated by about 1$^\circ$, which will be accomplished by a small maneuver immediately after their launch \cite{lator_cqg_2004,stanford_texas}. This separation would permit differential astrometric observations to an accuracy of $\sim 10^{-13}$ radians needed to significantly improve measurements of gravitational deflection of light in the solar gravity.

The schematic of the LATOR experiment is quite simple and is given in Figure \ref{fig:lator}. Two spacecraft are injected into a heliocentric solar orbit on the opposite side of the Sun from the Earth. The triangle in the figure has three independent quantities but three arms are monitored with laser metrology. Each spacecraft equipped with a laser ranging system that enable a measurement of the arms of the triangle formed by the two spacecraft and the ISS.   According to Euclidean rules this determines a specific angle at the interferometer; LATOR can measure this angle directly. In particular, the laser beams transmitted by each spacecraft are detected by a long baseline ($\sim$ 100 m) optical interferometer on the ISS. The actual angle measured at the interferometer is compared to the angle calculated using Euclidean rules and three side measurements; the difference is the non-Euclidean deflection signal (which varies in time during spacecraft passages), which contains the scientific information. 
This built-in redundant--geometry optical truss eliminates the need for drag-free spacecraft for high-accuracy navigation \cite{lator_cqg_2004}. 

The uniqueness of this mission comes with its geometrically redundant architecture that enables LATOR to measure the departure from Euclidean geometry ($\sim 8\times 10^{-6}$ rad) caused by the solar gravity field, to a very high accuracy.   This departure is shown as a difference between the calculated Euclidean value for an angle in the triangle and its value directly measured by the interferometer.  This discrepancy, which results from the curvature of the space-time around the Sun and can be computed for every alternative theory of gravity, constitutes LATOR's signal of interest. The precise  measurement of this departure constitutes the primary mission objective.

We now outline the basic elements of the LATOR trajectory.

\subsection{Spacecraft Trajectory: a 3:2 Earth Resonant Orbit}
\label{sec:3:2orbit}

\begin{figure*}[t!]
% \begin{center}
\hskip -10pt 
\begin{minipage}[b]{.46\linewidth}
\centering \psfig{figure=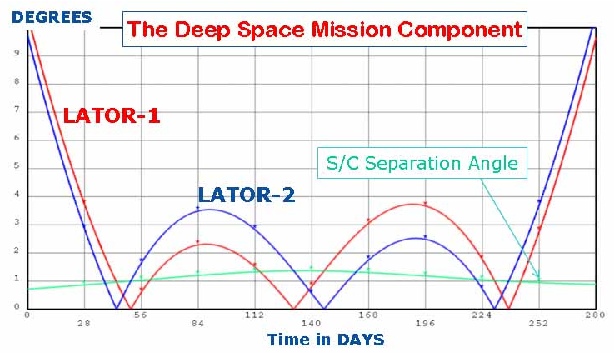,width=112mm}
\end{minipage}
\hfill  
\begin{minipage}[b]{.32\linewidth}
\centering 
\vbox{\psfig{figure=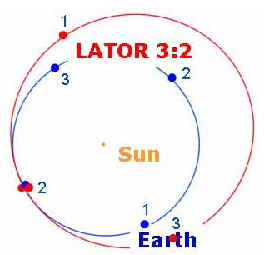,width=5.2cm}\\\vskip26pt}
\end{minipage}
\caption{Left: The Sun-Earth-Probe angle during the period of 3 occultations (two periodic curves) and the angular separation of the spacecraft as seen from the Earth (lower smooth line). Time shown is days from the moment when one of the spacecraft are at 10º distance from the Sun. Right: View from the North Ecliptic of the LATOR spacecraft in a 3:2 resonance. The epoch is taken near the first occultation.  
 \label{fig:lator_sep_angle2}}
% \end{center}
\vskip -5pt 
%\end{figure*}
%
%%*********************************
%%************
%\begin{figure*}[t!]
 \begin{center}
\noindent    
\psfig{figure=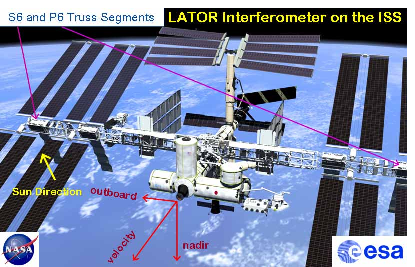,width=115mm}%,height=90mm}
%\end{center}
%\vskip -10pt 
%  \caption{Text.  
% \label{fig:lator_sep_angle}}
%\end{figure*} 
%
%%**************
%%************
%\begin{figure}[t!]
% \begin{center}
%\noindent    
\psfig{figure=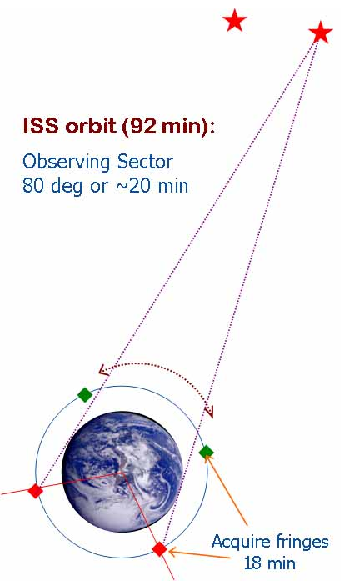,width=4.4cm}%,height=90mm}
%\end{center}
%\vskip -10pt 
  \caption{Left: Location of the LATOR interferometer on the ISS. To utilize the inherent ISS Sun-tracking capability, the LATOR optical packages will be located on the outboard truss segments P6 and S6 outwards. Right: Signal acquisition for each orbit of the ISS; variable baseline allows for solving fringe ambiguity.  
 \label{fig:iss_config}}
 \end{center}
\end{figure*}

To enable the primary objective, LATOR will place two spacecraft into a heliocentric orbit, to provide conditions for observing the spacecraft when they are behind the Sun as viewed from the ISS (see Figures~\ref{fig:iss_config},\ref{fig:lator_sep_angle2}).  With the help of the JPL Advanced Project Design Team (Team X), we recently conducted detailed mission design studies \cite{teamx}. In particular, we analyzed various trajectory options for the deep-space flight segment of LATOR, using both the Orbit Determination Program (ODP) and Satellite Orbit Analysis Program (SOAP)---the two standard JPL navigation software packages. 

%**************
%
%%**************
%\vskip 5pt 

An orbit with a 3:2 resonance with the Earth\footnote{The 3:2 resonance occurs when the Earth does 3 revolutions around the Sun while the spacecraft does exactly 2 revolutions on a 1.5 year orbit. The exact period of the orbit may vary slightly, $<$1\%, from a 3:2 resonance depending on the time of launch.} was found to uniquely satisfy the LATOR orbital requirements \cite{lator_cqg_2004,teamx}. 
For this orbit, 13 months after the launch, the spacecraft are within $\sim10^\circ$ of the Sun with first occultation occurring 15 months after launch \cite{lator_cqg_2004}.  At this point, LATOR is orbiting at a slower speed than the Earth, but as LATOR approaches its perihelion, its motion in the sky begins to reverse and the spacecraft is again occulted by the Sun 18 months after launch.  As the spacecraft slows down and moves out toward aphelion, its motion in the sky reverses again, and it is occulted by the Sun for the third and final time 21 months after launch.

Figure \ref{fig:lator_sep_angle2} shows the  trajectory  and the occultations in more details.  The figure on the right is the spacecraft position in the solar system showing the Earth's and LATOR's orbits (in the 3:2 resonance) relative to the Sun.  The epoch of this figure shows the spacecraft passing behind the Sun as viewed from the Earth.  The  figure on the left shows the trajectory when the spacecraft would be within 10$^\circ$ of the Sun as viewed from the Earth.  This period of 280 days will occur once every 3 years, provided the proper maneuvers are performed.  Two similar periodic curves give the Sun-Earth-Probe angles for the two spacecraft while the lower smooth curve gives their angular separation as seen from the Earth.

As a baseline design for the LATOR orbit, both spacecraft will be launched on the same launch vehicle. Almost immediately after the launch there will be a 30 m/s maneuver that separates the two spacecraft on their 3:2 Earth resonant orbits (see Figure \ref{fig:lator_sep_angle2}).  The sequence of events that occurs during each observation period will be initiated at the beginning of each orbit of the ISS. It assumed that bore sighting of the spacecraft attitude with the spacecraft transmitters and receivers have already been accomplished. This sequence of operations is focused on establishing the ISS to spacecraft link. The interspacecraft link is assumed to be continuously established after deployment, since the spacecraft never lose line of sight with one another (for more details consult Section~\ref{sec:operations}).

The 3:2 Earth resonant orbit provides an almost ideal trajectory for the LATOR mission, specifically i) it imposes no restrictions on the time of launch; ii) with a small propulsion maneuver after launch, it places the two LATOR spacecraft at the distance of less then 3.5$^\circ$ (or $\sim 14 ~R\odot$) for the entire duration of the experiment (or $\sim$8 months); iii) it provides three solar conjunctions even during the nominal mission lifetime of 22 months, all within a 7 month period; iv) at a cost of an extra maneuver, it offers a possibility of achieving  small orbital inclinations (to enable measurements at different solar latitudes); and, finally, v) it offers a very slow change in the Sun-Earth-Probe (SEP) angle of about $\sim R_\odot$ in 4 days. Furthermore, such an orbit provides three observing sessions during the initial 21 months after the launch, with the first session starting in 15 months \cite{lator_cqg_2004}. As such, this orbit represents a very attractive choice for LATOR. We intend to further study this 3:2 Earth resonant trajectory as the baseline option for the mission. 

\subsection{ISS-based Interferometer and Observing Sequence}
\label{sec:operations}

The interferometer on the ISS will be formed by the two optical assemblies with dimensions of approximately 0.6~m $\times$ 0.6~m $\times$ 0.6~m for each assembly (Figure~\ref{fig:iss_config}). The mass of each telescope assembly will be about 120~kg. The location of these packages on the ISS and their integration with the ISS's power, communication and attitude control system are given below:

\begin{itemize}
\item Two LATOR transponders will be physically located and integrated with the ISS infrastructure. The location will enable the straight-line separation between the two transponders of $\sim$100 m and will provide a clear line-of-site (LOS) path between the two transponders during the observation periods. Both transponder packages will have clear LOS to their corresponding heliocentric spacecraft during pre-defined measurement periods.   

\item The transponders will be physically located on the ISS structure to maximize the inherent ISS sun-tracking capability.  The transponders will need to point toward the Sun during each observing period.  By locating these payloads on the ISS outboard truss segments (P6 and S6 outward), a limited degree of automatic Sun-tracking capability is afforded by the $\alpha$-gimbals on the ISS.

\item The minimum unobstructed LOS time duration between each transponder on the ISS and the transponders and their respective spacecraft will be 58 minutes per the 92 minute orbit of the ISS.   
 
\item	The pointing error of each transponder to its corresponding spacecraft will be no greater than 1 $\mu$rad for control,  1 $\mu$rad for knowledge, with a stability of 0.1 $\mu$rad/sec, provided by a combination of the standard GPS link available on the ISS and $\mu$-g accelerometers. 

\end{itemize}

It is important to discuss the sequence of events that will lead to the signal acquisition and that occur during each observation period (i.e., every orbit of the ISS).  This sequence will be initiated at the beginning of the experiment period, after ISS emergence from the Earth's shadow (see Figure~\ref{fig:iss_config}). It assumes that boresighting of the spacecraft attitude with the spacecraft transmitters and receivers have already been accomplished. This sequence of operations is focused on establishing the ISS to spacecraft link. The interspacecraft link is assumed to be continuously established after final deployment (at $\sim15^\circ$ off the Sun), since the spacecraft never lose line of sight with one another. 

The laser beacon transmitter at the ISS is expanded to have a beam divergence of 30 arcsec in order to guarantee illumination of the LATOR spacecraft. After re-emerging from the Earth's shadow this beam is transmitted to the craft and reaches them in about 18 minutes. At this point, the LATOR spacecraft acquire the expanded laser beacon signal. In this mode, a signal-to-noise ratio (SNR) of 4 can be achieved with 30 seconds of integration. With an attitude knowledge of 10 arcsec and an array field of view of 30 arcsec no spiral search is necessary. Upon signal acquisition, the receiver mirror on the spacecraft will center the signal and use only the center quad array for pointing control. Transition from acquisition to tracking should take about 1 minute. Because of the weak uplink intensity, at this point, tracking of the ISS station is done at a very low bandwidth. The pointing information is fed-forward to the spacecraft transmitter pointing system and the transmitter is turned on. The signal is then re-transmitted down to the ISS with a light-travel time of 18 minutes.

Each interferometer station and laser beacon station searches for the spacecraft laser signal. The return is heterodyned by using an expanded bandwidth of 300~MHz. In this case, the solar background is the dominant source of noise, and an SNR of 5 is achieved with 1 second integration. Because of the small field of view of the array, a spiral search will take 30 seconds to cover a 30 arcsec field. Upon acquisition, the signal will be centered on the quad cell portion of the array and the local oscillator frequency locked to the spacecraft signal. The frequency band will then be narrowed to 5 kHz. In this regime, the solar background is no longer the dominant noise source and an SNR of 17.6 can be achieved in only 10 msec of integration. This will allow one to have a closed loop pointing bandwidth of greater than 100 Hz and be able to compensate for the tilt errors introduced by the atmosphere. The laser beacon transmitter will then narrow its beam to be diffraction limited ($\sim$1 arcsec) and to point toward the LATOR spacecraft. This completes the signal acquisition phase, and the entire architecture is in-lock and transmits scientific signal.  This procedure is re-established during each 92-minute orbit of the ISS.

We shall now consider the basic elements of the LATOR optical design. 

\subsection{General Principles of Optical Design}
\label{sec:optical_design}

%\subsection{Mission Optical Architecture}

\begin{figure*}[t!]
 \begin{center}
\noindent    
\epsfig{figure=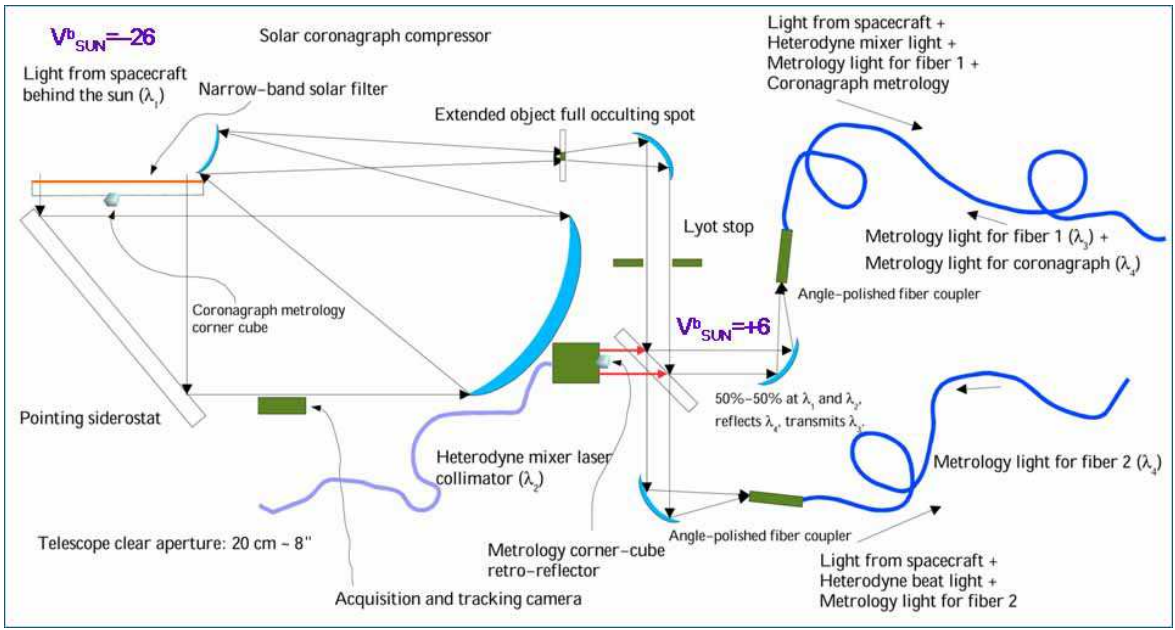,width=155mm}%,height=90mm}
\end{center}
\vskip -10pt 
  \caption{Basic elements of optical design for the LATOR interferometer: The laser light (together with the solar background) is going through a full aperture ($\sim20$cm) narrow band-pass filter with $\sim 10^{-4}$ suppression properties. The remaining light illuminates the baseline metrology corner cube and falls onto a steering flat mirror where it is reflected to an off-axis telescope with no central obscuration (needed for metrology). It then enters the solar coronograph compressor by first going through a 1/2 plane focal plane occulter and then coming to a Lyot stop. At the Lyot stop, the background solar light is reduced by a factor of $10^{6}$. The combination of a narrow band-pass filter and coronograph enables the solar luminosity reduction from $V=-26$ to $V=4$ (as measured at the ISS), thus enabling the LATOR precision observations.
\label{fig:optical_design}}
\vskip -5pt 
\end{figure*}

A single aperture of the interferometer on the ISS consists of three 20 cm diameter telescopes (see Figure \ref{fig:optical_design} for a conceptual design). One of the telescopes with a very narrow bandwidth laser line filter in front and with an InGaAs camera at its focal plane, sensitive to the 1064 nm laser light, serves as the acquisition telescope to locate the spacecraft near the Sun.

The second telescope emits the directing beacon to the spacecraft. Both spacecraft are served out of one telescope by a pair of piezo controlled mirrors placed on the focal plane. The properly collimated laser light ($\sim$10~W) is injected into the telescope focal plane and deflected in the right direction by the piezo-actuated mirrors. 

The third telescope is the laser light tracking interferometer input aperture, which can track both spacecraft at the same time. To eliminate beam walk on the critical elements of this telescope, two piezo-electric X-Y-Z stages are used to move two single-mode fiber tips on a spherical surface while maintaining focus and beam position on the fibers and other optics. Dithering at a few Hz is used to make the alignment to the fibers and the subsequent tracking of the two spacecraft completely automatic. The interferometric tracking telescopes are coupled together by a network of single-mode fibers whose relative length changes are measured internally by a heterodyne metrology system to an accuracy of less than 5~pm.

%************
%**************

The spacecraft  are identical in construction and contain a relatively high powered (1~W), stable (2~MHz per hour $\sim$500 Hz per second), small cavity fiber-amplified laser at 1064~nm. Three quarters of the power of this laser is pointed to the Earth through a 15~cm aperture telescope and its phase is tracked by the interferometer. With the available power and the beam divergence, there are enough photons to track the slowly drifting phase of the laser light. The remaining part of the laser power is diverted to another telescope, which points toward the other spacecraft. In addition to the two transmitting telescopes, each spacecraft has two receiving telescopes.  The receiving telescope, which points toward the area near the Sun, has laser line filters and a simple knife-edge coronagraph to suppress the Sun's light to 1 part in $10^4$ of the light level of the light received from the space station. The receiving telescope that points to the other spacecraft is free of the Sun light filter and the coronagraph.

In addition to the four telescopes they carry, the spacecraft also carry a tiny (2.5~cm) telescope with a CCD camera. This telescope is used to initially point the spacecraft directly toward the Sun so that their signal may be seen at the space station. One more of these small telescopes may also be installed at right angles to the first one, to determine the spacecraft attitude, using known, bright stars. The receiving telescope looking toward the other spacecraft may be used for this purpose part of the time, reducing hardware complexity. Star trackers with this construction were demonstrated many years ago and they are readily available. A small RF transponder with an omni-directional antenna is also included in the instrument package to track the spacecraft while they are on their way to assume the orbital position needed for the experiment. 

The LATOR experiment has a number of advantages over techniques that use radio waves to measure gravitational light deflection. Advances in optical communications technology, allow low bandwidth telecommunications with the LATOR spacecraft without having to deploy high gain radio antennae needed to communicate through the solar corona. The use of the monochromatic light enables the observation of the spacecraft almost at the limb of the Sun, as seen from the ISS. The use of narrowband filters, coronagraph optics and heterodyne detection will suppress background light to a level where the solar background is no longer the dominant noise source. In addition, the short wavelength allows much more efficient links with smaller apertures, thereby eliminating the need for a deployable antenna. Finally, the use of the ISS will allow conducting the test above the Earth's atmosphere---the major source of astrometric noise for any ground based interferometer. This fact justifies LATOR as a space mission.

\section{LATOR Optical Design}
\label{sec:interferometry}

In this section we consider the basic elements of the LATOR optical receiver system.  While we focus on the optics for the two spacecraft, the interferometer has essentially similar optical architecture.

\subsection{The LATOR Optical Receiver System}
 
%====================================

\begin{table*}[t!]
\begin{center}
\caption{Summary of design parameters for the LATOR optical receiver system.
\label{table:requirements}} \vskip 8pt
\begin{tabular}{rl} \hline\hline
Parameters/Requirements   & Value/Description \\\hline
 & \\[-10pt]
Aperture &  100 mm, unobstructed \\[3pt]
Wavelength & 1064 nm \\[3pt]
Narrow bandpass Filter & 2 nm FWHM over full aperture \\[3pt] 
Focal Planes & APD Data \& CCD Acquisition/Tracking \\[3pt]
APD Field of View & Airy disk field stop (pinhole) in front of APD\\[3pt]
APD Field Stop (pinhole) & Approximately 0.009 mm in diameter \\[3pt] 
APD Detector Size & TBD (a little larger than 0.009 mm)\\[3pt] 
CCD Field of View & 5 arc minutes \\[3pt] 
CCD Detector Size & 640 $\times$ 480 pixels (9.6 mm $\times$ 7.2 mm)\\[3pt]
CCD Detector Pixel Size & 15 $\mu$m\\[3pt] 
Beamsplitter Ratio (APD/CCD) & 90/10\\[3pt] 
Field Stop & `D'-shaped at primary mirror focus \\[3pt] 
Lyot Stop & Circular aperture located at telescope exit pupil\\[3pt] 
\hline\hline
\end{tabular} 
\end{center} 
\end{table*}
%==========================================================  

%************
\begin{figure*}[t!]
 \begin{center}
\noindent    
\psfig{figure=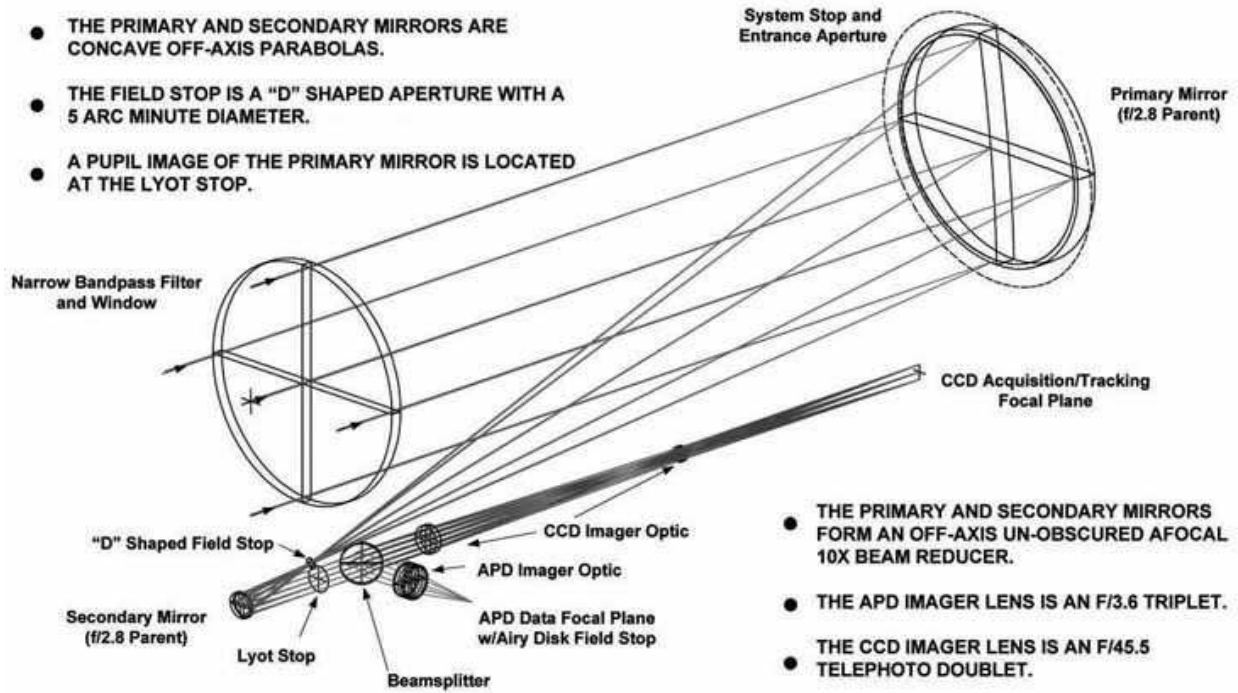,width=164mm}%,height=90mm}
\end{center}
\vskip -10pt 
  \caption{LATOR receiver optical system layout.  
 \label{fig:lator_receiver}
}
\end{figure*} 
%**************
%************
\begin{figure*}[t!]
 \begin{center}
\noindent    
\psfig{figure=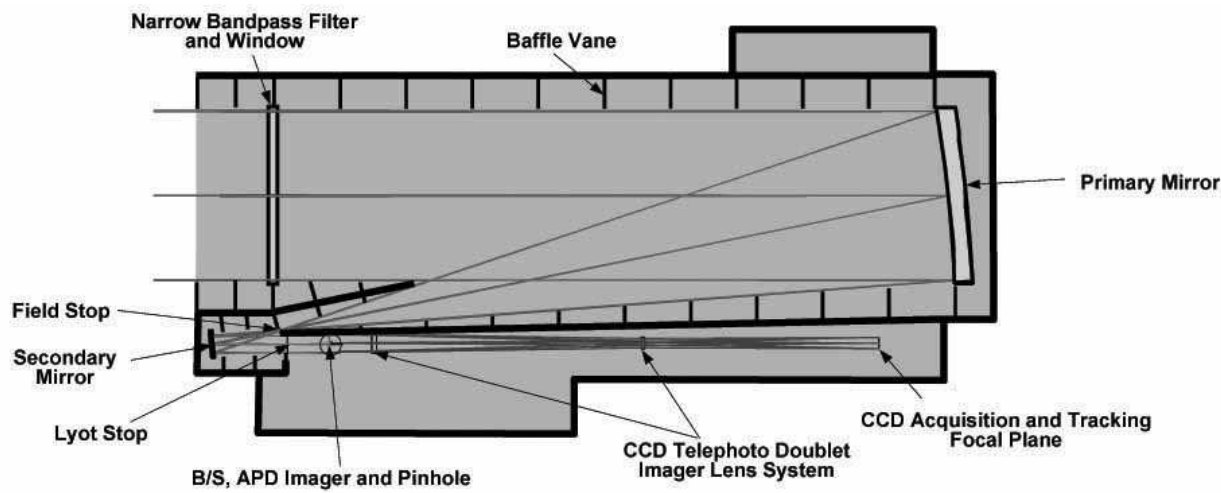,width=164mm}%,height=90mm}
\end{center}
\vskip -10pt 
  \caption{The LATOR preliminary baffle design.  
 \label{fig:lator_buffle} 
}
\end{figure*} 
%**************
%************
\begin{figure}[t!]
 \begin{center}
\noindent    
\psfig{figure=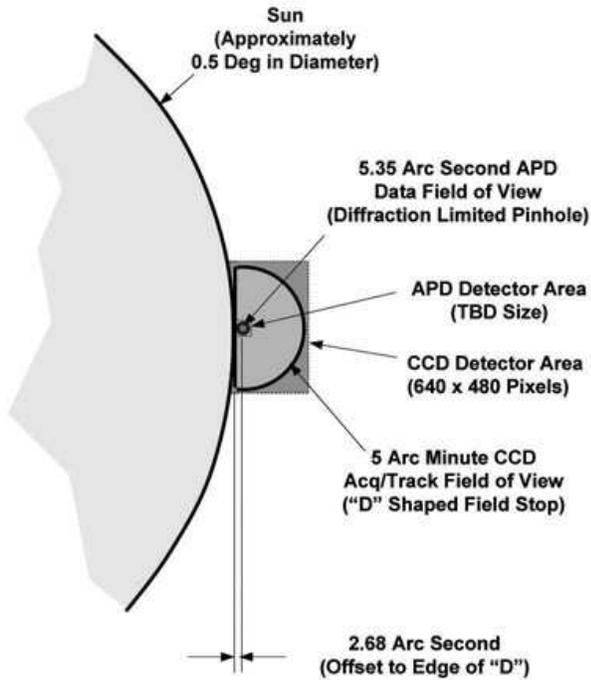,width=82mm}%,height=90mm}
\end{center}
\vskip -10pt 
  \caption{LATOR focal plane mapping (not to scale).  
 \label{fig:lator_focal}}
\end{figure} 
%**************

The LATOR 100 mm receiver optical system is a part of a proposed experiment. This system is located at each of two separate spacecraft placed on heliocentric orbits, as shown in Figure \ref{fig:lator}. The receiver optical system captures optical communication signals form a transmitter on the ISS, which orbits the Earth. To support the primary mission objective, this system must be able to receive the optical communication signal from the uplink system at the ISS that passes through the solar corona at the immediate proximity of the solar limb (at a distance of no more than 5 Airy disks).

Our recent analysis of the LATOR 100 mm receiver optical system successfully satisfied all the configuration and performance requirements (shown in Table \ref{table:requirements}) \cite{hellings_2005,stanford_texas,stanford_ijmpd}. We have also performed a conceptual design (see Figure \ref{fig:lator_receiver}), which was validated with a ray-trace analysis. The ray-trace performance of the designed instrument is diffraction limited in both the APD and CCD channels over the specified field of view at 1064 nm. The design incorporated the required field stop and Layot stop. A preliminary baffle design has been developed for controlling the stray light. 

The optical housing is estimated to have very accommodating dimensions; it measures (500 mm $\times$ 150 mm $\times$ 250 mm). The housing could be made even shorter by reducing the focal length of the primary and secondary mirrors, which may impose some fabrication difficulties. These design opportunities are being currently investigated. 

\subsubsection{Preliminary Baffle Design}

Figure \ref{fig:lator_buffle} shows the LATOR preliminary baffle design. The out-of-field solar radiation falls on the narrow band pass filter and primary mirror; the scattering from these optical surfaces puts some solar radiation into the FOV of the two focal planes. This imposes some requirements on the instrument design.  
Thus, the narrow band pass filter and primary mirror optical surfaces must be optically smooth to minimize narrow angle scattering. This may be difficult for the relatively steep parabolic aspheric primary mirror surface. However, the field stop will eliminate direct out-of-field solar radiation at the two focal planes, but it will not eliminate narrow angle scattering for the filter and primary mirror.  Finally, the Lyot stop will eliminate out-of-field diffracted solar radiation at the two focal planes. Additional baffle vanes may be needed at several places in the optical system. This design will be further investigated in series of trade-off studies by also focusing on the issue of stray light analysis. Figure \ref{fig:lator_focal} shows the design of the focal plane capping. The straight edge of the `D'-shaped CCD field stop is tangent to the limb of the Sun and it is also tangent to the edge of APD field stop. There is a 2.68 arcsecond offset between the straight edge and the concentric point for the circular edge of the CCD field stop.  The results of the analysis of APD and CCD channels point spread functions can be found in \cite{stanford_ijmpd,stanford_texas}. 

%************
\begin{figure*}[t!]
 \begin{center}
\noindent    
\psfig{figure=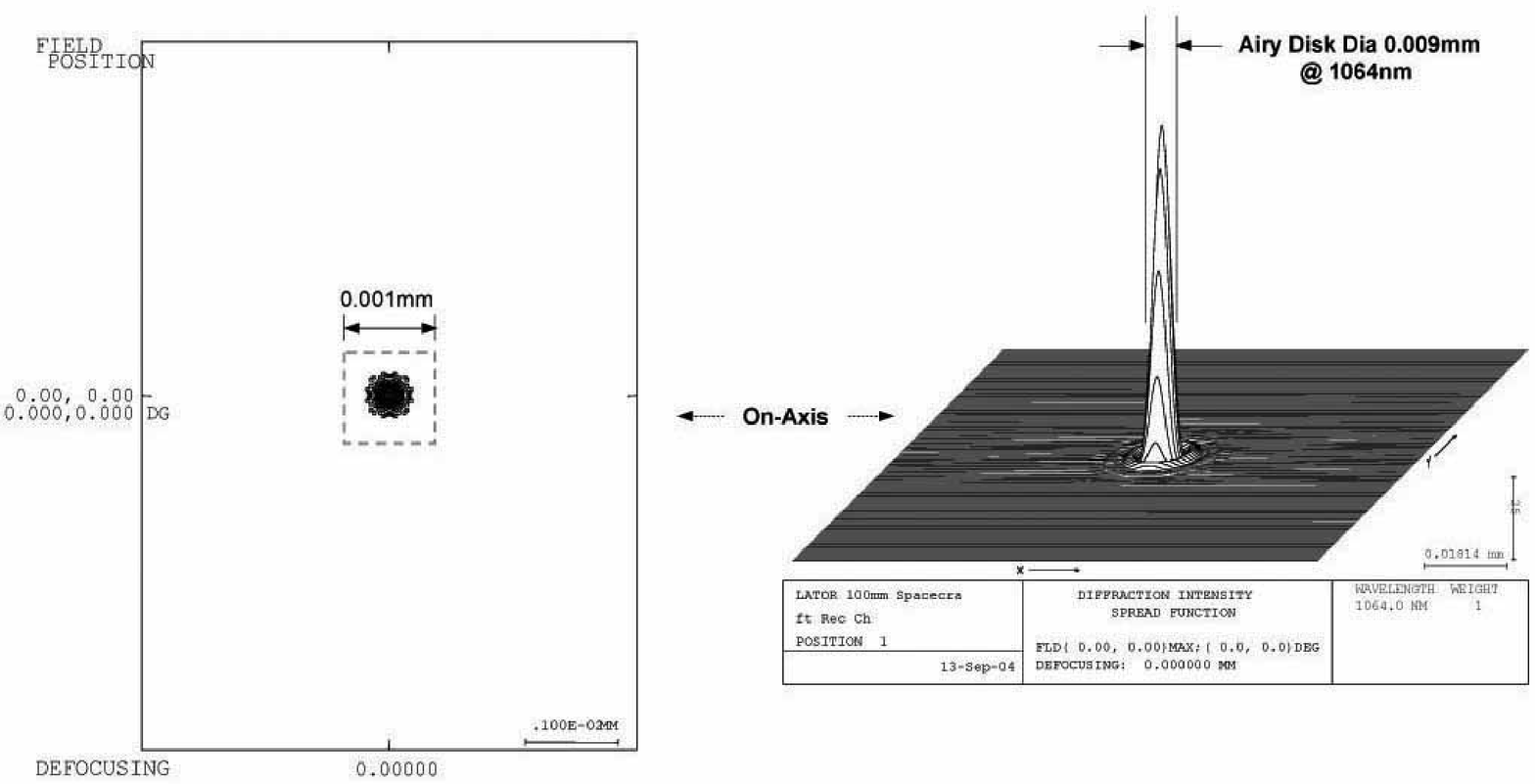,width=158mm}%,height=90mm}
\end{center}
\vskip -10pt 
  \caption{Simulated performance of the APD channel geometric (left) and diffraction (right) PSF.
 \label{fig:lator_apd}}
%\end{figure*} 
%%**************
%%************
%\begin{figure*}[t!]
 \begin{center}
\noindent    
\psfig{figure=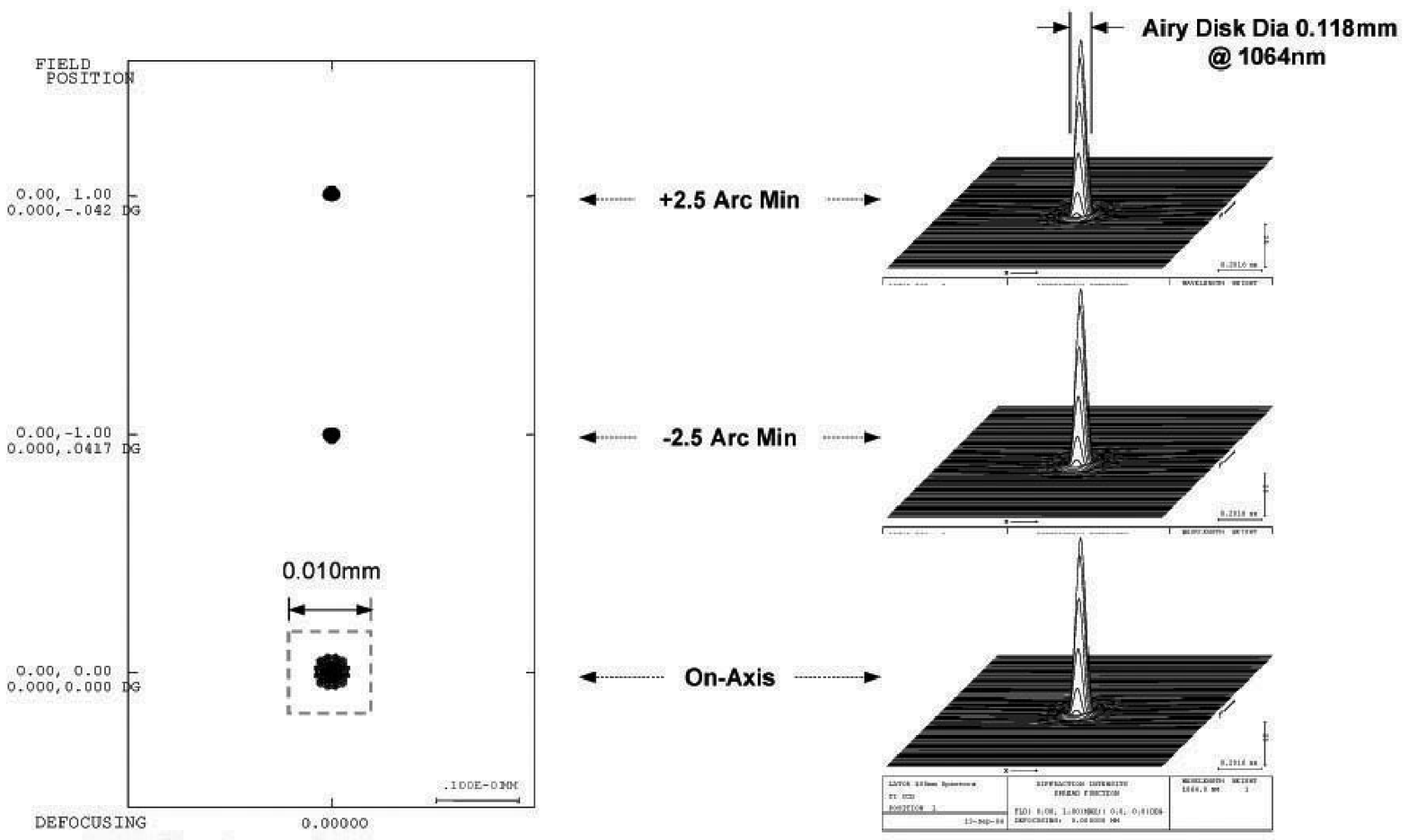,width=158mm}%,height=90mm}
\end{center}
\vskip -10pt 
  \caption{Simulated performance of the CCD channel geometric (left) and diffraction (right) PSF. 
 \label{fig:lator_ccd}}
\end{figure*} 
%**************

\subsubsection{Focal Plane Mapping}

Figure \ref{fig:lator_focal} shows the design of the focal plane capping. The straight edge of the `D'-shaped CCD field stop is tangent to the limb of the Sun and it is also tangent to the edge of APD field stop (pinhole). There is a 2.68 arcsecond offset between the straight edge and the concentric point for the circular edge of the CCD field stop (`D'-shaped aperture). In addition, the APD field of view and the CCD field of view circular edges are concentric with each other. Depending on the spacecraft orientation and pointing ability, the `D'-shaped CCD field stop aperture may need to be able to be rotated to bring the straight edge into a tangent position relative to the limb of the Sun. 
The results of the analysis of APD and CCD channels point spread functions (PSF) are shown in Figures \ref{fig:lator_apd} and \ref{fig:lator_ccd}.

\subsection{LATOR Coronograph}
\label{sec:coronograph}

\begin{figure}[h!]
 \begin{center}
\noindent  \vskip -5pt   
\psfig{figure=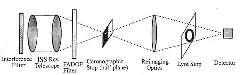,width=86mm}%,height=90mm}
\end{center}
\vskip -15pt 
  \caption{Schematics of the LATOR coronograph system. ~
 \label{fig:coronograph}} 
%{\bf text, text, text, tex}}
\end{figure}

In order to have adequate rejection of the solar background surrounding the laser uplink from Earth, the spacecraft optical system must include a coronagraph. Figure \ref{fig:coronograph} shows a schematic of the coronagraph. A 10 cm telescope forms an image on the chronographic stop. This stop consists of a knife-edge mask placed 6 arcseconds beyond the solar limb. The transmitted light is then reimaged onto a Lyot stop, which transmits 88\% of the incident intensity. Finally, the light is reimaged onto the tracking detector.
 
%%************
%**************
%%************
\begin{figure}[h!]
 \begin{center}
\noindent  \vskip 0pt   
\psfig{figure=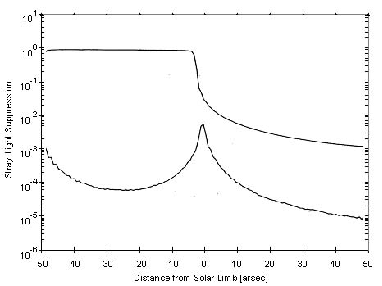,width=86mm}%,height=90mm}
\end{center}
\vskip -15pt 
  \caption{Results of coronograph performance simulation.
 \label{fig:coronograph_sim}}
\vskip -5pt 
\end{figure} 
%**************
 
The results of a simulated coronograph showing the stray light rejection as a function of the distance from the solar limb is shown in 
Figure~\ref{fig:coronograph_sim}. The solar surface has been approximated as a vertical edge extending along the entire length of a $256 \times 256$ array. The upper curve shows the stray light levels for an optical system without a coronograph. In this case, the flux from the solar surface has only been decreased by a factor of 100. With the coronograph, however, a further factor of 100 rejection can be achieved. In addition to decreasing the stray solar radiation, the coronograph will decrease the transmission of the laser signal by 78\% (for a signal 12 arcsec from limb) due to coronographic transmission and broadening of the point spread function. At these levels of solar rejection, it is possible for the spectral filter to reject enough starlight to acquire the laser beacon. It is interesting to note that without the coronograph, the stray light from the Sun, decreases proportionally to the distance from the limb. In contrast, with the use of the coronograph, the stray light decreases as the square of the distance from the limb.

\subsection{Factors Affecting SNR Analysis}

In conducting the signal-to-noise analysis we pay significant attention to several important factors. In particular, we estimate what fraction of the transmitted signal power captured by the 100 mm receiver aperture and analyze the effect of the Gaussian beam divergence (estimated at $\sim 7 ~\mu$rad) of the 200 mm transmit aperture on the ISS. Given the fact that the distance between the transmitter and receiver is on the order of 2 AU, the amount captured is about $2.3\times 10^{-10}$ of the transmitted power.  

We also consider the amount of solar disk radiation scattered into the two receiver focal planes. In particular, the surface contamination, coating defects, optical roughness and substrate defects could scatter as much as $1\times10^{-4}$ or more (possibly $1\times 10^{-3}$)  of the solar energy incident on the receive aperture into the field of view.   These issues are being considered in our current analysis. We also study the amount of the solar corona spectrum within the receive field of view that is not blocked by the narrow band pass filter.  The factors we consider is the filter's FWHM band-pass is 2 nm, the filter will have 4.0 optical density blocking outside the 2 nm filter band pass from the X-ray region of 1200 nm; the filter efficiency within the band pass will be about 35\%, and the detector is probably sensitive from 300 nm to 1200 nm.  

Additionally, we consider the amount of out-of-field solar radiation scattered into the focal plane by the optical housing. This issue needs to be investigated in a stray light analysis which can be used to optimize the baffle design to minimize the stray light at the focal plane.  Finally, we study the effectiveness of the baffle design in suppressing stray light at the focal plane. Thus, in addition to the stray light analysis, the effectiveness of the final baffle design should be verified by building an engineering model that can be tested for stray light.  

Our recent conceptual design and a CODEV  ray-trace analysis met all the configuration and performance requirements (shown in Table \ref{table:requirements}).  The ray-trace performance of the resulted instrument is diffraction limited in both the APD and CCD channels over the specified field of view at 1064 nm. The design incorporated the required field stop and Layot stop. A preliminary baffle design has been developed for controlling the stray light.  
In the near future, we plan to perform a stray light analysis which should be performed to optimize the baffle design and calculate the amount of stray light that could be present at each of the two focal planes.  This stray light analysis will take into account the optical smoothness of the band-pass filter and primary mirror surfaces. Narrow angle scattering may be a problem at the two focal planes in the filter and primary mirror are not optically very smooth and, thus, it requires a more detailed study. Finally, a rigorous signal-to-noise analysis will be performed to validate the power required to achieve a high signal-to-noise ratio in detecting received beam signal in the presence of the expected focal beam stray light predicted by the stray light analysis and the engineering model stray light tests.

The importance of this design is in the fact that it can be applied for many applications, thus, opening new ways for optical communication, accuracy navigational and fundamental physics experiments. This LATOR-related design experience motivates us to think about an architecture that may have a much border uses for the purposes of precision navigation and high data rate transmission and capable to operate at large interplanetary distance.
  
%*****************************
\section{Conclusions}  
\label{sec:conc}

The LATOR mission aims to carry out a test of the curvature of the solar system's gravity  field with an accuracy better than 1 part in 10$^{9}$. In spite of previous space missions exploiting radio waves for tracking the spacecraft, this mission manifests an actual breakthrough in the relativistic gravity experiments as it allow one to take full advantage of the optical techniques that recently became available.  
The LATOR experiment benefits from a number of advantages over techniques that use radio waves to study the light propagation in the solar vicinity.  The use of monochromatic light enables the observation of the spacecraft almost at the limb of the Sun, as seen from the ISS.  The use of narrowband filters, coronagraph optics, and heterodyne detection will suppress background light to a level where the solar background is no longer the dominant noise source.  The short wavelength allows much more efficient links with smaller apertures, thereby eliminating the need for a deployable antenna.  Advances in optical communications technology allow low bandwidth telecommunications with the LATOR spacecraft without having to deploy high gain radio antennae needed to communicate through the solar corona.  Finally, the use of the ISS not only makes the test affordable, but also allows conducting the experiment above the Earth's atmosphere---the major source of astrometric noise for any ground based interferometer.  This fact justifies the placement of LATOR's interferometer node in space. 

The LATOR mission will utilize several technology solutions that recently became available. In particular, signal acquisition on the solar background will be done with a full-aperture narrow band-pass filer and coronagraph. The issue of the extended structure vibrations of the ISS will be addressed by using $\mu$-g accelerometers. (The use of the accelerometers was first devised for SIM, but at the end their utilization was not needed. The Keck Interferometer uses accelerometers extensively.) Finally, the problem of monochromatic fringe ambiguity \cite{Shao_1995,Shao96,Yu94} is not an issue for LATOR. This is because the orbital motion of the ISS provides variable baseline projection that eliminates this problem for LATOR.  

The concept is technologically sound; the required technologies have been demonstrated as part of the international laser ranging activities and optical interferometry programs at JPL. The LATOR concept arose from several developments at NASA and JPL that initially enabled optical astrometry and metrology, and also led to developing expertize needed for the precision gravity experiments. Technology that has become available in the last several years such as low cost microspacecraft, medium power highly efficient solid state and fiber lasers, and the development of long range interferometric techniques make possible an unprecedented factor of 30,000 improvement in this test of general relativity. This mission is unique and is the natural next step in solar system gravity experiments that fully exploit modern technologies.

LATOR is envisaged as a partnership between NASA and ESA wherein both partners are essentially equal contributors, while focusing on different mission elements: NASA provides the deep space mission components and interferometer design, while building and servicing infrastructure on the ISS is an ESA contribution \cite{ESLAB2005_LATOR}. The NASA focus is on mission management, system engineering, software management, integration (both of the payload and the mission), the launch vehicle for the deep space component, and operations. The European focus is on interferometer components, the initial payload integration, optical assemblies and testing of the optics in a realistic ISS environment. The proposed arrangement would provide clean interfaces between familiar mission elements.

This mission may become a 21st century version of the Michelson-Morley experiment in the search for a  cosmologically evolved scalar field in the solar system. As such, LATOR will lead to very robust advances in the tests of fundamental physics: it could discover a violation or extension of general relativity, and/or reveal the presence of an additional long range interaction in the physical law. With this mission testing theory to several orders of magnitude higher precision, finding a violation of general relativity or discovering a new long range interaction could be one of this era's primary steps forward in fundamental physics. There are no analogs to the LATOR experiment; it is unique and is a natural culmination of solar system gravity experiments.

\section{Acknowledgements}

The work described here was carried out at the Jet Propulsion Laboratory, California Institute of Technology, under a contract with the National Aeronautics and Space Administration.

%\section{Appendix}
%
%\renewcommand{\theequation}{A.\arabic{equation}} %%% formulas will be numerates as (A.1), ...
%\setcounter{equation}{0}
%
%Appendix goes here.

%%**************************************


\begin{thebibliography}{99}

\bibitem%[\protect\astroncite{Turyshev, Shao, and Nordtvedt}{2004a}]
{lator_cqg_2004}  
Turyshev, S.\,G., Shao, M., and Nordtvedt, K.\,L., Jr.,\ 
``The Laser Astrometric Test of Relativity (LATOR) Mission,'' 
{\it Class. Quant. Grav. \bf 21}, 2773 (2004), gr-qc/0311020. 
%/ doi:10.1088/0264-9381/21/12/001

\bibitem%[\protect\astroncite{Nordtvedt}{1968b}]
{Ken_EqPrinciple68b}	
Nordtvedt, K.\,L., Jr.,  ``Equivalence Principle for Massive Bodies. II. Theory,'' {\it Phys. Rev. \bf 169},  1017 (1968).

\bibitem%[\protect\astroncite{Nordtvedt}{1968c}]
{Ken_LLR68}	
Nordtvedt, K.\,L., Jr., 
``Testing Relativity with Laser Ranging to the Moon,'' 
{\it Phys. Rev. \bf 170}, 1186 (1968).

\bibitem%[\protect\astroncite{Nordtvedt}{1987}]
{Ken_2PPN_87}	
Nordtvedt, K.\,L., Jr.,  ``Probing Gravity to the 2nd Post-Newtonian Order and to one part in 10$^7$ Using the Sun,'' 
{\it Astrophys. J. \bf320}, 871 (1987).

\bibitem%[\protect\astroncite{Will and Nordtvedt}{1972}]
{WillNordtvedt72}	
Will, C.\,M., Nordtvedt, K.\,L., Jr., ``Conservation Laws and Preferred Frames in Relativistic Gravity. I. Preferred-Frame Theories and an Extended PPN Formalism,'' Astrophys. J 177, 757 (1972).

\bibitem%[\protect\astroncite{Will}{1993}]
{Will_book93}	
Will, C.\,M., {\it Theory and Experiment in Gravitational Physics}, (Cambridge University Press, 1993).

\bibitem%[\protect\astroncite{Nordtvedt}{1996}]
{Ken_cqg96}	
Nordtvedt, K.\,L., Jr., ``Significance of `second-order' light propagation experiments in the solar system,'' 
{\it Class. Quant.  Grav. \bf13}, A11-A18 (1996).

\bibitem%[\protect\astroncite{Bertotti, Iess, and Tortora}{2003}]
{cassini_ber}
Bertotti,~B., Iess,~L., Tortora,~P., 
``A test of general relativity using radio links with the Cassini spacecraft,'' {\it Nature \bf 425}, 374  (2003).

\bibitem%[\protect\astroncite{Williams, Turyshev, and Murphy}{2004}]
{Williams_etal_2004}
Williams,~J.\,G., Turyshev,~S.\,G., and  Murphy,~T.\,W., Jr.,\ 
``Improving LLR Tests of Gravitational Theory,''
{\it Int. J. Mod. Phys. D\bf13},  567-582 (2004), gr-qc/0311021.

\bibitem%[\protect\astroncite{Williams, Turyshev, and Boggs}{2004}]
{LLR_beta_2004}
Williams,~J.\,G., Turyshev,~S.\,G., and  Boggs,~D.\,H.,\ 
``Progress in Lunar Laser Ranging Tests of Relativistic Gravity,''
{\it Phys. Rev. Lett. \bf93},  261101 (2004), gr-qc/0411113.

\bibitem%[\protect\astroncite{Damour and Esposito-Farese}{1996a}]
{Damour_EFarese96a}
Damour, T., Esposito-Farese, G., `` Testing gravity to second post-Newtonian order: a field-theory approach,''
{\it Phys. Rev. D\bf53}, 5541 (1996), gr-qc/9506063.

\bibitem%[\protect\astroncite{Damour and Esposito-Farese}{1996b}]
{Damour_EFarese96b}
Damour, T., Esposito-Farese, G., 
``Tensor-scalar gravity and binary pulsar experiments,''
{\it Phys. Rev. D\bf54}, 1474-1491 (1996), gr-qc/9602056.

\bibitem%[\protect\astroncite{Damour and Nordtvedt}{1993a}]
{Damour_Nordtvedt_93a} 
Damour, T., and Nordtvedt, K.\,L., Jr., 
``General Relativity as a Cosmological Attractor of Tensor Scalar Theories'', {\it Phys. Rev. Lett. \bf70}, 2217 (1993).

\bibitem%[\protect\astroncite{Damour and Nordtvedt}{1993b}]
{Damour_Nordtvedt_93b}	
Damour, T., and Nordtvedt, K.\,L., Jr., ``Tensor-scalar cosmological models and their relaxation toward general relativity,'' 
{\it Phys. Rev. D\bf48}, 3436  (1993).

\bibitem%[\protect\astroncite{Damour and Polyakov}{1994a}]
{DamourPolyakov94a}	
Damour, T., Polyakov, A.\,M., ``String Theory and Gravity,'' 
{\it Gen. Relativ. Gravit. \bf26}, 1171 (1994). 

\bibitem%[\protect\astroncite{Damour and Polyakov}{1994b}]
{DamourPolyakov94b}	
Damour, T., and Polyakov, A.,M., ``The string dilaton and a least coupling principle,'' {\it Nucl. Phys. B\bf423}, 532 (1994).

\bibitem%[\protect\astroncite{Damour, Piazza, and  Veneziano}{2002a}]
{DPV02a} 
Damour, T., Piazza, F., and  Veneziano, G.,
``Runaway dilaton and equivalence principle violations''
{\it Phys. Rev. Lett. \bf89}, 081601 (2002), gr-qc/0204094.

\bibitem%[\protect\astroncite{Damour, Piazza, and  Veneziano}{2002b}]
{DPV02b} 
Damour, T., Piazza, F., and  Veneziano, G.,
``Violations of the equivalence principle in a dilaton-runaway scenario,'' {\it Phys. Rev. D\bf66}, 046007 (2002), hep-th/0205111.

\bibitem%[\protect\astroncite{Damour and Esposito-Farese}{1993}]
{Damour_EFarese93} 
Damour, T., Esposito-Farese, G., `` Non-perturbative strong-field effects in tensor-scalar theories of gravitation,'' 
{\it Phys. Rev. Lett. \bf70}, 2220 (1993).

\bibitem%[\protect\astroncite{Bennett et~al.}{2003}]
{[4c]} 
Bennett, C.\,L., Halpern, M., Hinshaw, G., Jarosik, N., Kogut, A., Limon, M., Meyer, S.~S., Page, L., Spergel, D.~N., Tucker, G.~S., Wollack, E., Wright, E.~L., Barnes, C., Greason, M.~R., Hill, R.~S.,  Komatsu, E., Nolta, M.~R., Odegard, N., Peirs, H.~V., Verde, L., Weiland, J.~L., [i.e. WMAP Science Team],
``First Year Wilkinson Microwave Anisotropy Probe (WMAP) Observations: Preliminary Maps and Basic Results,'' 
{\it Astrophys. J. Suppl. \bf148}, 1 (2003), astro-ph/0302207.

\bibitem%[\protect\astroncite{de Bernardis et~al.}{2000}]
{deBernardis_CMB2000} 
de Bernardis, P., Ade, P.\,A.\,R., Bock, J.\,J., Bond, J.\,R., Borrill, J., Boscaleri, A., Coble, K., Crill, B.\,P., De Gasperis, G., Farese, P.\,C., Ferreira, P.\,G., Ganga, K., Giacometti, M., Hivon, E., Hristov, V.\,V., Iacoangeli, A., Jaffe, A.\,H., Lange, A.\,E., Martinis, L., Masi, S., Mason, P.\,V., Mauskopf, P.\,D., Melchiorri, A., Miglio, L., Montroy, T., Netterfield, C.\,B., Pascale, E., Piacentini, F., Pogosyan, D., Prunet, S., Rao, S., Romeo, G., Ruhl, J.\,E., Scaramuzzi, F., Sforna, D., Vittorio, N., ``A Flat Universe From High-Resolution Maps of the Cosmic Microwave Background Radiation,'' 
{\it Nature  \bf404}, 955-959 (2000).

\bibitem%[\protect\astroncite{Peacock et~al.}{2001}]
{Peacock_LargeScale01}	
Peacock, J.\,A., et al., ``A measurement of the cosmological mass density from clustering in the 2dF galaxy redshift survey,'' 
{\it Nature \bf410}, 169 (2001).

\bibitem%[\protect\astroncite{Perlmutter et~al.}{1999}]
{perlmutter99} 
Perlmutter, S., Aldering, G., Goldhaber, G., Knop, R.\,A., Nugent, P., Castro, P.\,G., Deustua, S., Fabbro, S., Goobar, A., Groom, D.\,E., Hook, I.\,M., Kim, A.\,G., Kim, M.\,Y., Lee, J.\,C., Nunes, N.\,J., Pain, R., Pennypacker, C.\,R., Quimby, R., Lidman, C., Ellis, R.\,S.,  Irwin, M., McMahon, R.\,G., Ruiz-Lapuente, P., Walton, N., Schaefer, B.,  Boyle, B.\,J., Filippenko, A.\,V.,  Matheson, T., Fruchter, A.\,S.,  Panagia, N., Newberg, H.\,J.\,M.,  Couch, W.\,J., [i.e. Supernova Cosmology Project Collaboration],
 ``Measurements of Omega and Lambda from 42 High-Redshift Supernovae,''  {\it Astrophys. J. \bf517}, 565-586  (1999), astro-ph/9812133.

\bibitem%[\protect\astroncite{Riess et~al.}{1999}]
{Riess_supernovae98}	
Riess,~A.\,G., Filippenko, A.\,V., Challis, P., Clocchiatti, A., Diercks, A., Garnavich, P.\,M., Gilliland, R.\,L., Hogan, C.\,J., Jha, S., Kirshner, R.\,., Leibundgut, B., Phillips, M.,M., Reiss, D., Schmidt, B.\,P., Schommer, R.\,A., Smith, R.\,C., Spyromilio, J., Stubbs, C., Suntzeff, N.\,B.; Tonry, J., [i.e., Supernova Search Team Collaboration], ``Observational Evidence from Supernovae for an Accelerating Universe and a Cosmological Constant,''   
{\it Astron. J. \bf116}, 1009-1038 (1998).

\bibitem%[\protect\astroncite{Tonry et~al.}{2003}]
{[3c]} 
Tonry, J.\,L.,  Schmidt, B.\,P.,  Barris, B.,  Candia, P., Challis, P.,  Clocchiatti, A.,  Coil, A.\,L., Filippenko, A.\,V.,  Garnavich, P.,  Hogan, C., Holland, S.\,T., Jha, S., Kirshner, R.\,P.,  Krisciunas, K.,  Leibundgut, B., Li, W.,  Matheson, T., Phillips, M.\,M., Riess, A.\,D.,  Schommer, R.,  Smith, R.\,C., Sollerman, J., Spyromilio, J., Stubbs, C.\,W.,  Suntzeff, N.\,B., 
``Cosmological Results from High-z Supernovae,''
{\it Astrophys. J. \bf594}, 1-24 (2003), astro-ph/0305008.

\bibitem%[\protect\astroncite{Halverson et~al.}{2002}]
{[6c]} 
Halverson, N.\,W., Leitch E.\,M., Pryke C., Kovac, J.,  Carlstrom, J.\,E., Holzapfel, W.\,L., Dragovan, M., Cartwright, J.\,K., Mason, B.\,S., Padin, S., Pearson, T.\,J., Shepherd, M.\,C.,  Readhead, A.\,C.\,S.,
 ``DASI First Results: A Measurement of the Cosmic Microwave Background Angular Power Spectrum,'' {\it Astrophys. J. \bf568}, 38 (2002), astro-ph/0104489. 

\bibitem%[\protect\astroncite{Netterfield et~al.}{2002}]
{[5c]} 
Netterfield, C.,\,B., Ade, P.\,A.\,R., Bock, J.\,J., Bond, J.\,R., Borrill, J., Boscaleri, A.,  Coble, K., Contaldi, C.\,R., Crill, B.\,P., de Bernardis, P., Farese, P.,  Ganga, K., Giacometti, M., Hivon, E., Hristov, V.\,V., Iacoangeli, A., Jaffe, A.\,H., Jones, W.\,C., Lange, A.\,E., Martinis, L., Masi, S., Mason, P., Mauskopf, P.\,D., Melchiorri, A., Montroy, T., Pascale, E.,  Piacentini, F., Pogosyan, D., Pongetti, F., Prunet, S., Romeo, G.,  Ruhl, J.E., Scaramuzzi, F.,  
[i.e. Boomerang Collaboration], ``A measurement by BOOMERANG of multiple peaks in the angular power spectrum of the cosmic microwave background,''  {\it Astrophys. J. \bf571}, 604 (2002), astro-ph/0104460. 

\bibitem%[\protect\astroncite{Carroll}{2001}]
{Carroll_01} 
Carroll~S.\,M.,  ``The Cosmological Constant,'' 
{\it Living Rev. Rel. \bf4}, 1 (2001), astro-ph/0004075.

\bibitem%[\protect\astroncite{Peebles  and Ratra}{2003}]
{PeeblesRatra03} 
Peebles, P.\,J.\,E., and Ratra, B., ``The Cosmological Constant and Dark Energy,'' {\it Rev. Mod. Phys. \bf75}, 559-606 (2003), astro-ph/0207347.

\bibitem%[\protect\astroncite{Carroll et~al.}{2004}]
{[carroll]} 
Carroll,~S.\,M., Duvvuri,~V., Trodden,~M., and Turner,~M., ``Is Cosmic Speed-Up Due to New Gravitational Physics?''
{\it Phys. Rev. D\bf70}, 043528 (2004), astro-ph/0306438.

\bibitem%[\protect\astroncite{Carroll, Hoffman, and Trodden}{2003}]
{Carroll_HT_03} 
Carroll, S.\,M., Hoffman, M., and Trodden, M., ``Can the dark energy equation-of-state parameter $w$ be less than $-1$?'' 
{\it Phys. Rev. D\bf68}, 023509 (2003), astro-ph/0301273.

\bibitem%[\protect\astroncite{Nordtvedt}{2003}]
{Ken_LLR_PPNprobe03}	
Nordtvedt, K.\,L., Jr.,  ``Lunar Laser Ranging - A Comprehensive Probe of Post-Newtonian Gravity'', (2003), gr-qc/0301024.

\bibitem%[\protect\astroncite{Epstein and Shapiro}{1980}]
{epstein_shapiro_80}	
Epstein, R., Shapiro, I.\,I., 
``Post-post-Newtonian deflection of light by the Sun,'' 
{\it Phys. Rev. D\bf22}, 2947 (1980).

\bibitem%[\protect\astroncite{Fischbach and  Freeman}{1980}]
{FishbachFreeman80}	
Fischbach, E., and  Freeman, B.\,S., ``Second-order contribution to the gravitational deflection of light,'' {\it Phys. Rev. D\bf22}, 2950 (1980). 

\bibitem%[\protect\astroncite{Richter and Matzner}{1982a}]
{RichterMatzner82a}	
Richter, G.\,W., and Matzner, R.\,A., ``2nd-order contributions to relativistic time delay in the parametrized post-Newtonian formalism,'' {\it Phys. Rev. D\bf26},  1219 (1982).

\bibitem%[\protect\astroncite{Richter and Matzner}{1982b}]
{RichterMatzner82b}	
Richter,~G.\,W., Matzner,~R.\,A.,
``Second-order contributions to gravitational deflection of light in the parametrized post-Newtonian formalism. II. Photon orbits and deflections in three dimensions,'' {\it Phys. Rev. D\bf26}, 2549 (1982).

\bibitem%[\protect\astroncite{Richter and Matzner}{1983}]
{RichterMatzner83}
Richter,~G.\,W., Matzner,~R.\,A., ``Second-order contributions to relativistic time delay in the parametrized post-Newtonian formalism,'' {\it Phys. Rev. D\bf28}, 3007-3012 (1983).

\bibitem%[\protect\astroncite{Plowman and Hellings}{2005}]
{hellings_2005}	
Plowman,~J.\,E., Hellings,~R.\,W., 
``LATOR Covariance Analysis,'' (2005), gr-qc/0505064.

\bibitem%[\protect\astroncite{Nordtvedt}{2005}]
{Ken_lator05}	
Nordtvedt, K.\,L., Jr.,   
``Covariance analysis studies for LATOR mission,'' In proc. ``359th WE-Heraeus Seminar: ``Lasers, Clock, and Drag-Free: Technologies for Future Exploration in Space and Gravity Tests,'' University of Bremen, ZARM, Bremen, Germany, 30 May - 1 June 2005, to be published, Springer-Werlag (2006). 

\bibitem%[\protect\astroncite{Turyshev, Shao, and Nordtvedt}{2004c}]
{stanford_texas} 
Turyshev, S.\,G., Shao, M., and Nordtvedt, K.\,L., Jr.,\ 
``Optical Design for the Laser Astrometric Test of Relativity'',   
in Proc. ``The XXII Texas Symposium on Relativistic Astrophysics,'' Stanford University, December 13-17, 2004,  
eConf C041213 \#0306 (2004), gr-qc/0502113.

\bibitem%[\protect\astroncite{Gerber et~al.}{2003}]
{teamx} 
Gerber, A., et al., ``LATOR 2003 Mission Analysis,''  JPL Advanced Project Design Team (Team X) Report \#X-618 (2003).

\bibitem%[\protect\astroncite{Turyshev, Shao, and Nordtvedt}{2004b}]
{stanford_ijmpd} 
Turyshev, S.\,G., Shao, M., and Nordtvedt, K.\,L., Jr.,\ 
``Experimental Design for the LATOR Mission,''   
{\it Int. J. Mod. Phys. D\bf13}, 2035 (2004), gr-qc/0410044.

 
\bibitem%[\protect\astroncite{Shao}{1995}]
{Shao_1995}
Shao, M., ''Prospects for Ground Based Interferometric Astrometry,'' {\it Astrophys. Space Sci. \bf223}, 119 (1995).

\bibitem%[\protect\astroncite{Shao et~al.}{1996}]
{Shao96}
Shao, M., Yu, G., G\"ursel, Y., Hellings, R.,  et~al. ``Laser Astrometric Test of Relativity (LATOR),'' 
JPL Internal Technical Memorandum,  (1996).

\bibitem%[\protect\astroncite{Yu et~al.}{1994}]
{Yu94} 
Yu, J., Shao, M., Gursel Y., and Hellings R.,  ``LATOR Mission,''
{\it SPIE Publ. \bf2200}, 325 (1994).

\bibitem%[\protect\astroncite{Turyshev et~al.}{2005}]
{ESLAB2005_LATOR} 
Turyshev, S.\,G., Dittus, H., L\"ammerzahl, C., Theil, S., Ertmer, W., Rasel, E., Foerstner, R., Johann, U., Klioner, S., Soffel, M., Dachwald, B., Seboldt, W., Perlick, V., Sandford, M.\,C.\,W., Bingham, R., Kent, B., Sumner, T.\,J., Bertolami, O., P\'aramos, J., Christophe, B., Foulon, B., Touboul, P., Bouyer, P., Damour, T., Salomon, C., Reynaud, S., Brillet, A.,  Bondu, F., Mangin, J.-F., Samain, E., Erd, C., Grenouilleau, J.\,C., Izzo, D., Rathke, A., Asmar, S.\,W., Colavita, M., G\"ursel, Y., Hemmati, H., Shao, M., Williams, J.\,G., Nordtvedt, K.\,L., Jr., Degnan, J., Plowman, J.\,E., Hellings, R., Murphy, T.\,W., Jr., ``Fundamental Physics with the Laser Astrometric Test Of Relativity,'' In proceesings of ``2005 ESLAB Symposium "Trends in Space Science and Cosmic Vision 2020,''  ESA/ESTEC, Noordwijk, The Netherlands, 19 April 2005, to be published, (2005), gr-qc/0506104.

\end{thebibliography}
\end{document}